\documentclass[11pt,epsf,letterpaper]{article}%
\usepackage{setspace}
\usepackage{color}
\usepackage{amsmath}
\usepackage{amsfonts}
\usepackage{verbatim}
\usepackage{amssymb}
\usepackage{graphicx}%
\providecommand{\U}[1]{\protect\rule{.1in}{.1in}}
\begin{document}

\title{Non-linear superposition law and Skyrme crystals}
\author{Fabrizio Canfora$^{1,2}$\\$^{1}$\textit{Centro de Estudios Cient\'{\i}ficos (CECS), Casilla 1469,
Valdivia, Chile.}\\$^{2}$\textit{Universidad Andres Bello, Av. Republica 440, Santiago, Chile.}\\{\small canfora@cecs.cl}}
\maketitle

\begin{abstract}
Exact configurations of the four-dimensional Skyrme model are presented. The
static configurations have
the profile which behaves as a kink and, consequently, the corresponding energy-momentum tensor describes a domain-wall. Furthermore,
a class of exact time-periodic Skyrmions is discovered. Within such class, it
is possible to disclose a remarkable phenomenon which is a genuine effect of
the Skyrme term. For a special value of the frequency the Skyrmions admit a
non-linear superposition principle. One can combine two or more exact
"elementary" Skyrmions (which may depend in a non-trivial way on all the
space-like coordinates) into a new exact composite Skyrmion. Due to such
superposition law, despite the explicit presence of non-linear effects in the
energy-momentum tensor, the interaction energy between the elementary
Skyrmions can be computed exactly. The relations with the appearance of Skyrme
crystals is discussed.

\end{abstract}

\section{Introduction}

The Skyrme theory is one of the most important model of theoretical physics
due to its wide range of applications. Skyrme \cite{skyrme} introduced his
famous term into the action of the non-linear sigma model to allow the
existence of static soliton solutions with finite energy, called
\textit{Skyrmions} (see~ \cite{multis2}\ \cite{manton}
\cite{susy} for detailed pedagogical reviews). A further important property of
the Skyrme theory is that excitations around Skyrme solitons may represent
Fermionic degrees of freedom (see the detailed analysis in \cite{bala2} and
references therein) suitable to describe nucleons (nice examples are
\cite{bala0} \cite{bala1} \cite{ANW} \cite{guada}). Besides the great
importance of Skyrmions in particles, nuclear physics and astrophysics (see,
for instance, \cite{astroSkyrme}\ and references therein), many intriguing
observations such as \cite{useful1}, \cite{useful2}, \cite{useful3},
\cite{useful4}, \cite{useful5}, \cite{useful6}, \cite{useful7.1},
\cite{useful7} and \cite{useful8} show that Skyrmions play a very important
role also in Bose-Einstein condensates, nematic liquids, multi-ferric
materials, chiral magnets and condensed matter physics in general. However, in condensed matter systems (such as the ones
considered in \cite{useful7.1}), Skyrmions are stabilized by specific
interactions imposed by handedness of the underlying structure (for details,
see \cite{useful7} \cite{useful8}). This mechanism is absent in the original
Skyrme model and, consequently, as it is well known it is extremely difficult
to construct analytical configurations in that case. Another issue that makes
the task to construct non-trivial analytical configurations of the original
Skyrme model particularly difficult is the fact that, unlike what happens for
instance in the case of monopoles and instantons in Yang-Mills-Higgs theory
(see, for instance,\ \cite{manton} \cite{susy} \cite{moduli}), the Skyrme-BPS bound on the energy (which was derived by Skyrme himself well before the modern formulation of solitons theory became available) cannot be
saturated for non-trivial spherically symmetric Skyrmions. Nevertheless, in
\cite{skyrmonopole} it has been shown that the Skyrme model supports knotted
excitations (but they are not known analytically) as it also happens in
Yang-Mills theory (see, in particular, \cite{knotYM}).

Among the many open problems of the Skyrme theory (in which exact
configurations are very rare\footnote{One of the very few examples (which is
not of the type analyzed in the present paper) of exact solutions of the
Skyrme field equation is in \cite{example1}. Due to the non-linearities of the
theory, explicit solutions are available only under severe simplifying
assumptions.}) a fundamental theoretical challenge is to construct exact
configurations in which the non-linear effects are explicitly present. This
would shed considerable light on the peculiar interactions of the Skyrme
model. A further theoretical problem is to explain how, in a non-linear and
non-integrable theory such as the Skyrme model in four dimensions, nice
crystal-like structures (see \cite{sut1} and references therein) are able to
appear. These beautiful configurations, which have been constructed
numerically, look almost like non-trivial super-positions of "elementary
solutions" but, at a first glance, it appears to be impossible to accommodate
both non-linear effects and (any sort of) superposition law.

One often adopts suitable ansatz to make the field equations more tractable.
Until recently, the only ansatz able to reduce the complexity Skyrme field
equations is the hedgehog ansatz for spherically symmetric systems and its
\textit{rational map} generalization\footnote{In both cases (the usual
spherical hedgehog and the rational map ansatz), it is necessary that the
profile of the hedgehog depends on the radius and, consequently, it is not
easy to depart from spherical symmetry.} in \cite{rational}. However, the
field equations are still unsolvable in these cases, since the corresponding
Skyrme-BPS bound cannot be saturated.

Many modifications of the Skyrme theory have been proposed (see, for instance,
\cite{simple1} \cite{simple2} \cite{simple3} \cite{simple4} \cite{simple5}
\cite{simple6} and references therein) which change the theory in such a way
that suitable bounds on the energy which can be saturated become available. All these models have
many interesting mathematical features and their importance goes beyond the
simplification of the original Skyrme theory. However, all such modifications
spoil the original motivations of the Skyrme theory whose form is compelling
both from phenomenological and theoretical points of view.

In the present paper, it will be shown that using the generalized hedgehog
ansatz introduced in \cite{canfora} \cite{canfora2} (which allows to study
systems without spherical symmetry) one can construct exact configurations of
the four-dimensional Skyrme theory in its original form which may be static or periodic in time (but
other topologies can be considered as well). Within the time-periodic
configurations, a remarkable resonance phenomenon emerges: for a special value
of the frequency, a non-linear superposition law appears which allows to
combine two or more "elementary" Skyrmions\footnote{It is worth to emphasize
here that, often, in the literature the term "Skyrmion" is used only to
describe configurations with non-vanishing winding number. In the following,
this term will be used in a broader sense to denote exact solutions of the original
four-dimensional Skyrme theory in which non-linear effects are explicitly
present.}
(whose profile depends in a non-trivial way on all the space-like coordinates)
into a new exact composite Skyrmion. In the present case, despite the explicit
presence of non-linear effects, the interaction energy between the moduli of
$N$ elementary Skyrmions \textit{can be computed exactly for any }$N$. The
relations with the appearance of Skyrme crystals is an intriguing
property of the present construction.

This paper is organized as follows: in the second section, the generalized
hedgehog will be described and an exact kink of the Skyrme system will be
constructed. In the third section, exact time-periodic Skyrmions will be
analyzed and it will be shown how a non-linear superposition law arises. In
the fourth section it will be discussed a mapping from periodic Skyrmions in
which the profile is static and the internal orientation is time-periodic to
periodic Skyrmions in which the profile depends on time while the internal
orientation is periodic in space. In the fifth section, some conclusions will
be drawn.

\section{Generalized hedgehog and an exact kink}

The action of the $SU(2)$ Skyrme system in four dimensional flat (but
topologically non-trivial) space-times is
\begin{align}
S_{\mathrm{Skyrme}}  &  =\frac{K}{2}\int d^{4}x\sqrt{-g}\mathrm{Tr}\left(
\frac{1}{2}R^{\mu}R_{\mu}+\frac{\lambda}{16}F_{\mu\nu}F^{\mu\nu}\right)
\ ,\ \ \ K>0\ ,\ \ \ \lambda>0\ ,\label{skyrmaction}\\
R_{\mu}  &  :=U^{-1}\nabla_{\mu}U=R_{\mu}^{i}t_{i}\ ,\ \ F_{\mu\nu}:=\left[
R_{\mu},R_{\nu}\right]  \ ,\ \ \ \hbar=1\ ,\ \ \ c=1\ ,\nonumber
\end{align}
where the Planck constant and the speed of light have been set to $1$, the
coupling constants $K$ and $\lambda$ are fixed by comparison with experimental
data and the $t^{i}$ are the basis of the $SU(2)$ generators $t^{i}$ (where
the Latin index $i$ corresponds to the group index).

A mass term can also be included: however it is well known (see \cite{sut2}
and references therein) that when the mass term is non-vanishing, the theory
is considered in 1+1 dimensions and the group is $U(1)$ kinks of Sine-Gordon
type saturating suitable BPS bounds appear. Here it will be shown that this is
true even \textit{without a mass term} in 3+1 dimensions and with the $SU(2)$
internal symmetry group. Furthermore, the original action\footnote{It is worth
mentioning that Skyrme original idea was precisely that the nucleon's mass is
of mesonic origin and that the meson mass originates from the nucleon
coupling.} in Eq. (\ref{skyrmaction}) without the mass term (besides to be a
quite good approximation for Pions physics when the Pions mass can be assumed
to be small) is extremely useful in applications in condensed matter physics
(see, for instance \cite{useful2} and \cite{useful4}).

The Skyrme field equations are
\begin{equation}
\nabla^{\mu}R_{\mu}+\frac{\lambda}{4}\nabla^{\mu}[R^{\nu},F_{\mu\nu}]=0.
\label{nonlinearsigma1}%
\end{equation}

The following standard parametrization of the $SU(2)$-valued scalar $U(x^{\mu
})$ will be adopted%
\begin{equation}
U^{\pm1}(x^{\mu})=Y^{0}(x^{\mu})\mathbf{1}\pm Y^{i}(x^{\mu})t_{i}%
\ ,\ \ \left(  Y^{0}\right)  ^{2}+Y^{i}Y_{i}=1\ . \label{standard1}%
\end{equation}
The generalized hedgehog ansatz \cite{canfora} \cite{canfora2} reads
\begin{equation}
Y^{0}=\cos\alpha\ ,\ \ Y^{i}=\widehat{n}^{i}\sin\alpha\ , \label{genehedge1}%
\end{equation}%
\begin{equation}
\widehat{n}^{1}=\cos\Theta\ ,\ \ \ \widehat{n}^{2}=\sin\Theta\ ,\ \ \ \widehat
{n}^{3}=0\ ,\ \left(  \nabla_{\mu}\alpha\right)  \left(  \nabla^{\mu}%
\Theta\right)  =0\ , \label{genehedge2}%
\end{equation}
where $\alpha$ is the profile of the hedgehog while the function $\Theta$,
through the internal vector $\widehat{n}^{i}$, describes its orientational
degrees of freedom in the internal $SU(2)$ space.

As it has been discussed in \cite{canfora} \cite{canfora2} a very convenient
choice is to take $\Theta$\ as a linear function of the coordinates in such a
way that, when the metric is flat, one gets%
\[
\left(  \nabla_{\mu}\Theta\right)  \left(  \nabla^{\mu}\Theta\right)  =\left(
\nabla\Theta\right)  ^{2}=const\neq0\ ,
\]
since the system becomes rather trivial when $\left(  \nabla\Theta\right)
^{2}$ vanishes (in which case the field equation for $\alpha$ linearizes and,
mainly, the energy-momentum tensor becomes quadratic in $\alpha$ so that the
non-linear effects disappear). 
Thus, the full Skyrme field equations Eq. (\ref{nonlinearsigma1}) reduce to the
following single scalar non-linear partial differential equation for the
Skyrmion profile $\alpha$:%
\begin{equation}
0=\left(  1+\lambda\left(  \nabla\Theta\right)  ^{2}\sin^{2}\alpha\right)
\square\alpha+\frac{\lambda\left(  \nabla\Theta\right)  ^{2}\left(
\nabla\alpha\right)  ^{2}}{2}\sin\left(  2\alpha\right)  -\frac{\left(
\nabla\Theta\right)  ^{2}}{2}\sin\left(  2\alpha\right)  =0\ ,\ \ \square
=\nabla_{\mu}\nabla^{\mu}\ , \label{fieldequ}%
\end{equation}
while the $t-t$ component of the energy-momentum (which represents the
energy-density) reads%
\begin{align}
T_{tt}  &  =K\left\{  \left(  \nabla_{t}\alpha\right)  \left(  \nabla
_{t}\alpha\right)  +\sin^{2}\alpha\left(  \nabla_{t}\Theta\right)  \left(
\nabla_{t}\Theta\right)  -\frac{g_{tt}}{2}\left[  \left(  \nabla\alpha\right)
^{2}+\sin^{2}\alpha\left(  \nabla\Theta\right)  ^{2}\right]  +\right.
\label{toogen}\\
&  \left.  \lambda\sin^{2}\alpha\left(  \left(  \nabla\Theta\right)
^{2}\left(  \nabla_{t}\alpha\right)  \left(  \nabla_{t}\alpha\right)  +\left(
\nabla\alpha\right)  ^{2}\left(  \nabla_{t}\Theta\right)  \left(  \nabla
_{t}\Theta\right)  \right)  -\frac{\lambda g_{tt}}{2}\sin^{2}\alpha\left(
\left(  \nabla\Theta\right)  ^{2}\left(  \nabla\alpha\right)  ^{2}\right)
\right\}  \ .\nonumber
\end{align}

It is worth emphasizing the following point. When one replaces a suitable ansatz into the equations of motion the informations about the correct boundary conditions are lost. In such cases, the correct boundary conditions for the Skyrmion profile $\alpha$ can be derived analysing, for instance, the energy-density.
In the present case, the global minima of the energy-density in Eq.
(\ref{toogen}) are located at $\alpha=n\pi$. Hence, the two most natural possibilities are to require periodic boundary conditions for $\alpha$ (in which cases one is effectively considering the theory defined on a torus with finite volume\footnote{This situation is especially relevant for the applications of Skyrmions in condensed matter physics in which the relevant dynamics happen in finite domains.}) or to require that $\alpha$ approaches $n\pi$ at the boundaries of the region one is analysing. Both cases will be considered in the following.

As an interesting static example, one can consider the following ansatz for the Skyrme field:
\begin{align}
\alpha &  =\alpha\left(  x\right)  \ ,\Theta=\Theta(y,z)=\frac{1}{l}(n_{2}y+n_{3}z)
\ ,\ \ \ n_{2}\ ,\ n_{3}\ \in%
\mathbb{Z}
\ ,\label{ans1}\\
(\nabla\alpha)^{2}  &  =\left(  \alpha^{\prime}\right)  ^{2}\ ,\ \ \ \alpha
^{\prime}=\frac{d\alpha}{dx}\ ,\ (\nabla\Theta)^{2}=L^{-2}=(\left(
n_{2}\right)  ^{2}+\left(  n_{3}\right)  ^{2})l^{-2}\ , \label{ans2}%
\end{align}
where $L$ and $l$ are constants with dimension of length (so that both $\Theta$ and the product $\lambda L^{-2}$ are adimensional) while $x$, $y$ and $z$ are the spatial coordinates. The choice in Eq. (\ref{ans1}) corresponds to the internal vector $\widehat{n}^{i}$ periodic in the $y$ and $z$ spatial directions while the profile $\alpha$ only depends on the spatial coordinate $x$.
The full Skyrme field equations Eq. (\ref{fieldequ}) and the energy-density in Eq.
(\ref{toogen}) reduce to
\begin{equation}
0=\alpha^{\prime\prime}-\frac{L^{-2}}{2}\sin(2\alpha)+\lambda L^{-2}\left[
\alpha^{\prime\prime}\sin^{2}\alpha+\frac{(\alpha^{\prime})^{2}}{2}%
\sin(2\alpha)\right]  \ , \label{eomm}%
\end{equation}%
\begin{equation}
T_{tt}=\frac{K}{2}\left[  (\alpha^{\prime})^{2}+L^{-2}\sin^{2}\alpha+\lambda
L^{-2}(\alpha^{\prime})^{2}\sin^{2}\alpha\right]  \ . \label{t00}%
\end{equation}

In this case, the correct boundary conditions correspond to require that $\alpha$ at $x=\pm\infty$ approaches to absolute minima of the energy density located at $n\pi$. It is worth mentioning that such boundary condition makes sense not only from the energetic point of view. The reason is that, when $\alpha$ is equal to  $n\pi$, the Skyrmion configurations defined in Eqs. (\ref{standard1}), (\ref{genehedge1}) and (\ref{genehedge2}) reduces to $\pm\mathbf{1}$ and it is customary to require that the Skyrme configuration reduces to (plus or minus) the identity at the boundaries of the domains one is interested in.

At a first glance, it is a very difficult task to find exact solutions of Eq. (\ref{eomm}). However, one can observe that any solution of the following first order differential equation for $\alpha$ %
\begin{equation}
\alpha^{\prime}=\pm\frac{L^{-1}\sin\alpha}{\left(  F(\alpha)\right)  ^{1/2}%
}\ , \label{BPS6}%
\end{equation}
\begin{equation}
\ F(\alpha)=1+\lambda
L^{-2}\sin^{2}\alpha\ , \label{BPS1}%
\end{equation}
is automatically a solution of Eq. (\ref{eomm}) as well. Indeed, if $\alpha$ satisfies
Eq. (\ref{BPS6}) then one can deduce (deriving both sides of Eq. (\ref{BPS6}))
the following expression for the second derivative of $\alpha$:%
\begin{equation}
\alpha^{\prime\prime}=\frac{L^{-2}}{2}\left\{  \frac{\sin\left(  2\alpha
\right)  }{1+\lambda L^{-2}\sin^{2}\alpha}-\frac{\lambda L^{-2}\sin\left(
2\alpha\right)  \sin^{2}\alpha}{\left(  1+\lambda L^{-2}\sin^{2}\alpha\right)
^{2}}\right\}  \ . \label{BPS7}%
\end{equation}
Then, replacing the expressions in Eqs. (\ref{BPS6}) and (\ref{BPS7}) for
$\alpha^{\prime}$ and $\alpha^{\prime\prime}$ into Eq. (\ref{eomm}) one can
easily see that Eq. (\ref{eomm}) is satisfied identically.
By writing Eq. (\ref{BPS6}) in the form of a one-dimensional Newton equation,
one can show there exist kink-like solutions which interpolate between
neighbouring minima of the energy-density in Eq. (\ref{t00}) located at
$n\pi$. Due to the fact that the profile $\alpha$ interpolates between neighbouring minima, this type of solutions cannot be deformed continuously to the trivial solutions $\alpha=n\pi$. The obvious reason is that, to perform such deformation, one would need an infinite amount of energy per unit of area in the $y-z$ plane. In particular, the above configuration represents actually a domain wall (for a detailed pedagogical review see \cite{vilenkin}). Indeed, the full energy-momentum is%

\begin{align}
T_{\mu\nu}  &  =K\left\{  \left(  \nabla_{\mu}\alpha\right)  \left(
\nabla_{\nu}\alpha\right)  +\sin^{2}\alpha\left(  \nabla_{\mu}\Theta\right)
\left(  \nabla_{\nu}\Theta\right)  -\frac{\eta_{\mu\nu}}{2}\left[  \left(
\nabla\alpha\right)  ^{2}+\sin^{2}\alpha\left(  \nabla\Theta\right)
^{2}\right]  +\right. \label{tooogen}\\
&  \left.  \lambda\sin^{2}\alpha\left(  \left(  \nabla\Theta\right)
^{2}\left(  \nabla_{\mu}\alpha\right)  \left(  \nabla_{\nu}\alpha\right)
+\left(  \nabla\alpha\right)  ^{2}\left(  \nabla_{\mu}\Theta\right)  \left(
\nabla_{\nu}\Theta\right)  \right)  -\frac{\lambda \eta_{\mu\nu}}{2}\sin
^{2}\alpha\left(  \left(  \nabla\Theta\right)  ^{2}\left(  \nabla
\alpha\right)  ^{2}\right)  \right\}  \ ,\nonumber
\end{align}
$\eta_{\mu\nu}$ being the flat Minkowski metric. 
Since Eq. (\ref{BPS6}) implies that the profile $\alpha$ is a monotone function interpolating between neighbouring minima of the energy density, one can see that the full energy-momentum tensor is different from zero only in a narrow region around $x=0$ (for any $y$ and $z$) and it approaches exponentially fast to zero far from the $x=0$ plane. Consequently, as domain walls are defined by energy-momentum tensors which are localized around a plane and which approach to zero exponentially fast far from the plane (see \cite{vilenkin}), the above energy-momentum tensor corresponding to the Skyrmion profile $\alpha$ satisfying Eq. (\ref{BPS6}) represents a domain wall.
To the best of author's knowledge, this is the first exact domain wall in the four-dimensional Skyrme theory: due to the great importance of domain walls in many applications (see \cite{vilenkin}) this result is quite interesting.
It is also worth noting that, obviously, the total energy of a domain wall is divergent because of the integration along the transverse $y$ and $z$ directions. In fact, the relevant quantity in these cases is the total energy per unit of area in the $y$ and $z$ directions (which will be denoted as $E_A$) which reads:

\begin{equation}
E_A=\frac{K}{2}\int_{-\infty}^{+\infty}\left[  (\alpha^{\prime})^{2}+L^{-2}\sin^{2}\alpha+\lambda
L^{-2}(\alpha^{\prime})^{2}\sin^{2}\alpha\right]  \ dx \rightarrow \nonumber %
\end{equation}
\begin{equation}
E_A=\frac{K}{2}\int_{-\infty}^{+\infty}(1+\lambda
L^{-2}sin^{2}\alpha)[  (\alpha^{\prime})^{2}+\frac{L^{-2}}{(1+\lambda
L^{-2}sin^{2}\alpha)} \sin^{2}\alpha]dx. \nonumber%
\end{equation}
One can rewrite the above equation as
\begin{equation}
E_A=\frac{K}{2}\int_{-\infty}^{+\infty}(1+\lambda
L^{-2}sin^{2}\alpha)[  (\alpha^{\prime})^{2}+\frac{L^{-2}}{(1+\lambda
L^{-2}sin^{2}\alpha)} \sin^{2}\alpha+ \nonumber%
\end{equation}
\begin{equation}
+2\alpha^{\prime}[\frac{L^{-2}}{(1+\lambda
L^{-2}sin^{2}\alpha)} \sin^{2}\alpha]^{1/2}-2\alpha^{\prime}[\frac{L^{-2}}{(1+\lambda
L^{-2}sin^{2}\alpha)} \sin^{2}\alpha]^{1/2}]   dx \rightarrow \nonumber%
\end{equation}
\begin{equation}
E_A=\frac{K}{2}\int_{-\infty}^{+\infty}(1+\lambda
L^{-2}sin^{2}\alpha)[  \alpha^{\prime}\pm(\frac{1}{(1+\lambda
L^{-2}sin^{2}\alpha)})^{1/2}L^{-1}\sin\alpha]^{2}   dx \mp Q \label{boundterm0}%
\end{equation}

\begin{equation}
Q=K\int_{-\infty}^{+\infty}[L^{-1}\sin\alpha(1+\lambda
L^{-2}sin^{2}\alpha)^{1/2}]  \alpha^{\prime}   dx  \label{boundterm}%
\end{equation}

Since the expression for $Q$ in Eq. (\ref{boundterm}) is a total derivative and that, due to the boundary conditions, $\alpha$ approaches at $x=\pm\infty$ to absolute minima of the energy density itself, the expression for $Q$ is convergent. Moreover, when the profile $\alpha$ satisfies Eq. (\ref{BPS6}), the expression for $E_A$ in Eq. (\ref{boundterm0}) reduces to $Q$ and so it is convergent as well. Therefore, the above configuration has the expected physical behaviour of a domain wall.

It is worth emphasizing here that the above example shows that, within Skyrme theory, there are
many more non-trivial topological effects than the ones encoded in the winding
number\footnote{The fact that the winding number by itself is not enough to
describe all the subtle topological properties of the Skyrme model can be
understood, for instance, following the analysis of \cite{skyrmonopole}.}.

\section{Time-periodic Skyrmions and non-linear superposition law}

Now, a class of exact time-periodic solutions (denoted as "periodic
Skyrmions") will be presented. Such class corresponds to the following
alternative choices of the profile of the hedgehog $\alpha$ and of
the function $\Theta$:%

\begin{equation}
\alpha=\alpha(x,y,z)  \ ,\ (\nabla\alpha)^{2}%
=(\partial_{x}\alpha)^{2} +(\partial_{y}\alpha)^{2}+(\partial_{z}\alpha)^{2}  , \label{ans1bis}%
\end{equation}
\
\begin{equation}
\Theta=\omega t\ ,\ \ \omega\ \in%
\mathbb{R}
\ ,\ \ (\nabla\Theta)^{2}=-\omega^{2}\ , \label{ans3bis}%
\end{equation}
where the metric is the flat Minkowski metric in Cartesian coordinates, $x$, $y$ and $z$ are the spatial coordinates while boundary conditions along the spatial direction will be
specified below. The ansatz in Eq. (\ref{ans3bis}) describes Skyrmions with a
profile $\alpha$ which depends on all the space-like coordinates\footnote{In
the cases in which $\alpha$ depends on only one space-like variable, one can
find exact solutions following the same construction described in the previous
section.}. On the other hand, the internal vector $\widehat{n}^{i}$ which
describes the orientation of the Skyrmion in the internal $SU(2)$ space
oscillates in time with frequency $\omega$ between the\ first and the second
generators of the $SU(2)$ algebra: the description of this dynamical
situation would be impossible with the usual spherical hedgehog ansatz.

With the above choice of $\alpha$ and $\Theta$, the Skyrme field equations Eq. (\ref{nonlinearsigma1}) reduce to (see  \cite{canfora} \cite{canfora2}) the following scalar elliptic
non-linear partial differential equation for the Skyrmion profile%
\begin{equation}
0=\left(  1-\lambda\omega^{2}\sin^{2}\alpha\right)  \triangle\alpha
+\frac{\omega^{2}}{2}\sin(2\alpha)-\frac{\lambda\omega^{2}\sin(2\alpha)}%
{2}(\nabla\alpha)^{2}\ , \label{eom2}%
\end{equation}
where $\triangle$ is the flat three-dimensional Laplacian.

In the following, periodic boundary conditions for the profile $\alpha$ will be considered. This corresponds effectively to compactify the space to a three-dimensional torus of finite volume. Skyrme theory in finite domains can be quite relevant in applications in condensed matter physics as well (as the references cited in the introduction shows). How the analysis changes with different boundary conditions will be mentioned at the end of this section.

One can observe that in terms of the following function $H(\alpha)$ of the
profile $\alpha$%
\begin{equation}
H\left(  \alpha\right)  =\int^{\alpha}\sqrt{1-\lambda\omega^{2}\sin^{2}%
s}ds\ \Rightarrow\frac{\triangle H}{\sqrt{1-\lambda\omega^{2}\sin^{2}\alpha}%
}=\left[  \triangle\alpha-\frac{\lambda\omega^{2}}{2}\frac{\sin(2\alpha
)(\nabla\alpha)^{2}}{1-\lambda\omega^{2}\sin^{2}\alpha}\right]  \ ,
\label{ans7}%
\end{equation}
Eq. (\ref{eom2}) can be written as%
\begin{equation}
\triangle H+\frac{\omega^{2}\sin(2\alpha)}{2\sqrt{1-\lambda\omega^{2}\sin
^{2}\alpha}}=0\ , \label{eom2bis}%
\end{equation}
where one should express $\alpha$ in terms of $H$ inverting the elliptic
integral in Eq. (\ref{ans7}). A very surprising phenomenon is now apparent: if%
\begin{equation}
\omega=\omega^{\ast}=\frac{1}{\sqrt{\lambda}}\ , \label{freres}%
\end{equation}
then Eq. (\ref{eom2}) with the change of variable in Eq. (\ref{ans7}) reduces
to the following (\textit{linear!}) Helmholtz equation%
\begin{equation}
\triangle H+\frac{1}{\lambda}H=0\ ,\ \ \ \alpha=\arcsin H\ , \label{eom3bis}%
\end{equation}
which, as it will be discussed below, allows to define an exact\textit{
non-linear superposition law}. The energy-density (defined in Eq.
(\ref{toogen})) in terms of $H$ becomes%
\begin{equation}
T_{tt}=K\left\{  \frac{1}{2\lambda}H^{2}+\frac{1}{2}\left(  \frac{1+H^{2}%
}{1-H^{2}}\right)  \left(  \nabla H\right)  ^{2}\right\}  \ . \label{combt00}%
\end{equation}

The explicit presence of non-linear effects in the energy-momentum tensor in
Eq. (\ref{combt00}) despite the fact that $H$ satisfies the linear Helmholtz
equation is related to the breaking of the homogeneous scaling
symmetry. Unlike what happens in free field theories, the energy-density does not scale
homogeneously under the rescaling
\[
H\rightarrow\rho H\ ,\ \ \rho\in%
\mathbb{R}
\ .
\]
In particular, by definition (see Eqs. (\ref{eom3bis}) and (\ref{genehedge1}%
)), $\left\vert H\right\vert $\ cannot be larger than $1$ and, moreover, from
the energetic point of view, it may be very "expensive" for $\left\vert
H\right\vert $ to get close to $1$ as it is clear from Eq. (\ref{combt00}).
Therefore, one can multiply a given solution $H_{(0)}$ for a constant $\rho$
whenever the (absolute value of the) new solution $\rho H_{(0)}$ of the
Helmholtz equation does not exceed $1$. 

The non-analytic dependence of the energy density in Eq. (\ref{combt00}) on
the Skyrme coupling $\lambda$ clearly shows both the non-perturbative nature
of the present effect and the fact that it is closely related to the Skyrme
term\footnote{Namely, it disappears when $\lambda\rightarrow0$ as it is clear
from Eq. (\ref{ans7}).}. Here only the cases in which Eq. (\ref{eom3bis}) has
a unique solution up to integration constants (which play the role of moduli
of the Skyrmions) will be considered since they allow a more transparent
physical interpretation (but more general situations can be analyzed as well).
The periodic Skyrmions corresponding such unique solution will be denoted as
\textit{elementary Skyrmions}.

Let us consider the case in which the soliton profiles depend on three
space-like coordinates $x$, $y$ and $z$. Periodic boundary conditions (with
periods $2\pi L_{i}$) in the spatial directions will be considered. Let
\begin{align}
H_{i}  &  =H\left(  \overrightarrow{x}-\overrightarrow{x}_{i}\right)
=A_{i}\left(  \sin\mu_{1}\left(  x-x_{i}\right)  \sin\mu_{2}\left(
y-y_{i}\right)  \sin\mu_{3}\left(  z-z_{i}\right)  \right)  \ , \label{multi1}%
\\
\frac{1}{\lambda}  &  =\sum_{i=1}^{3}\mu_{i}^{2}\ ,\ \ \ \mu_{i}=\frac
{1}{L_{i}}\ ,\ \overrightarrow{x}_{i}=\left(  x_{i},y_{i},z_{i}\right)
\ ,\ \ \alpha_{i}=\arcsin H_{i}\ , \label{multi2}%
\end{align}
where $H\left(  \overrightarrow{x}-\overrightarrow{x}_{i}\right)  $ is the
solution of Eq. (\ref{eom3bis}). As one can check in Eq. (\ref{combt00}), the
positions of the peaks in the energy density in Eq. (\ref{combt00})
corresponding to the elementary Skyrmions $\alpha_{i}=\arcsin H_{i}$ are
determined by $\overrightarrow{x}_{i}$ which therefore plays the role of the
moduli of $\alpha_{i}$ since $\overrightarrow{x}_{i}$\ identifies the
"position" of the elementary Skyrmion. On the other hand, the overall constant
$A_{i}$ strictly speaking does not represent a moduli of the elementary
Skyrmion since, when one replaces the expression in Eq. (\ref{multi1}) into
Eq. (\ref{combt00}), one can see that the total energy depends on $A_{i}$
while it does not depend on $\overrightarrow{x}_{i}$.

The most natural way to define a continuous composition of $N$ elementary
Skyrmions $\alpha_{i}=\arcsin H_{i}$ with moduli $\overrightarrow{x}_{i}%
$\ (defined in Eqs. (\ref{multi1}) and (\ref{multi2})) with the property that,
when all the $H_{i}$ are small, the profile of the sum reduces to the sum of
the profiles is%
\begin{align}
\alpha_{1+2+..+N}  &  =\arcsin\left(  H_{1}+H_{2}+..+H_{N}\right)
\ ,\ \ if\ \ \ \left\vert H_{1}+H_{2}..+H_{N}\right\vert \leq
1\ ,\label{second1}\\
\alpha_{1+2+..+N}  &  =\frac{\pi}{2}\ ,\ \ if\ \ \ H_{1}+H_{2}..+H_{N}%
>1\ ,\label{second2}\\
\alpha_{1+2+..+N}  &  =-\frac{\pi}{2}\ ,\ \ if\ \ \ H_{1}+H_{2}..+H_{N}<-1\ ,
\label{second3}%
\end{align}
where one must take into account that $\arcsin x$ is only defined when
$\left\vert x\right\vert \leq1$. At a first glance, in the cases in which $\left\vert H_{1}+H_{2}+..\right\vert >1$, discontinuities in the first derivatives of the composite Skyrmion can appear. In fact, it is unlikely that such non-smooth solutions can survive since they are very expensive energetically (since the corresponding gradient would be unbounded). Hence, the non-linear superposition of elementary Skyrmions is only allowed when they satisfy  $\left\vert H_{1}+H_{2}+..\right\vert <1$ otherwise it is not energetically convenient to combine the elementary Skyrmions into the composite Skyrmion.

As far as the appearance of crystals is concerned, from Eqs. (\ref{toogen})
and (\ref{combt00}) one can see that the energy-density of the composition
$\alpha_{1+2+..+N}$ of $N$ elementary Skyrmions (whose integral represents the
interaction energy between the $N$ elementary Skyrmions and can be computed in
principle for any $N$) depends in a complicated way on the moduli
$\overrightarrow{x}_{i}$. However, one can observe in Eqs. (\ref{toogen}) and
(\ref{combt00}) that, in the expression of the energy-density of the
composition $\alpha_{1+2+..+N}$, in order to find configurations of the
$\overrightarrow{x}_{i}$\ which are favorable energetically, one should
minimize with respect to the moduli $\overrightarrow{x}_{i}$ quadratic sums of
the following type
\[
\Psi=\left(  \sum_{i=1}^{N}H\left(  \overrightarrow{x}-\overrightarrow{x}%
_{i}\right)  \right)  ^{2}\ .
\]
In all the cases in which $H\left(  \overrightarrow{x}\right)  $ involves
trigonometric functions, the theory of interference in optics\footnote{Namely,
if one considers sums such as $\sum_{j}^{N-1}A\exp(i\xi_{j})$ one can show
that it vanishes when $\xi_{j}-\xi_{j-1}$ is equal to $2\pi/N$ for any $j$. In
the present case, the difference between the arguments in the summands\ is
related to the distance between peaks of neighboring Skyrmions.} can be
applied to minimize sums of the type appearing in the above equation since one
can interpret $\Psi$ as the interference of many elementary waves. Hence,
placements in which the $\overrightarrow{x}_{i}$ follow patterns of negative
interference are always favorable energetically (although other local minima
of the total energy appear as well). In the cases in which $H\left(
\overrightarrow{x}\right)  $ involves a different basis of functions (such as
the Bessel functions which naturally appear when analyzing the Helmholtz
equation in unbounded domains) the known results in optics cannot be applied
directly, but it is reasonable to expect that also in those cases the
$\overrightarrow{x}_{i}$ follow patterns associated to "negative interference
of Bessel functions".

It is worth emphasizing here that the present composite Skyrmions do not
correspond to the Skyrme crystals already known numerically (see for instance
\cite{sut1} \cite{manton}) since, in the latter case, the internal $SU(2)$
orientation depends non-trivially on space-like coordinates while, in the
present case, it depends non-trivially on time. However, the above result is
quite remarkable since, to the best of author's knowledge, it is the first
example constructed in a non-linear non-integrable four dimensional theory in
which it is possible to accommodate both explicit non-linear effects in the
energy-momentum tensor and a superposition law of "elementary" solutions.
Hence, the present results shed new light on the non-linear interactions which
are responsible for the appearance of composite structures such as the ones in
\cite{sut1} \cite{manton}.

It is natural to wonder whether the intriguing picture described above is a
"coincidence" which can only appear when the frequency of the periodic
Skyrmion satisfies Eq. (\ref{freres}). In fact, standard results in
perturbation theory (see the detailed analysis in \cite{berger}) ensure that
there exist a non-trivial left neighborhood of $\omega=1/\sqrt{\lambda}$ in
which the above construction provides a good approximation of the
multi-Skyrmions solutions. To see this, let us consider $\lambda\omega
^{2}=1-\eta$ where $0<\eta\ll1$. In this case, one can treat the term
proportional to $\eta$ in the Skyrme field equation Eq. (\ref{eom2}) as a
perturbation and $\eta$\ as the perturbation parameter. Well known techniques
in the theory of non-linear partial differential equations can be applied to
the present case (see, in particular, page 206, chapters 4.5B and 4.5C of
\cite{berger}). The kind of results that one gets in this way is that, given a
solution $\alpha_{(0)}$ of Eq. (\ref{eom2}) with $\eta=0$, there exist a
unique solution $\alpha_{(\eta)}$ of Eq. (\ref{eom2}) with $\eta>0$ and small
such that $\alpha_{(\eta)}\underset{\eta\rightarrow0}{\rightarrow}\alpha
_{(0)}$ uniformly apart from a narrow region close to the boundary. The
uniqueness result implies that, for small enough $\eta$, the composite
periodic Skyrmions described above when $\eta=0$ will exist also when $\eta$
is positive and small enough. Of course, they will be slightly distorted and
their energies will acquire corrections of order $\eta$, but the broad
theoretical picture will remain the same.

Besides periodic boundary conditions, there are two more interesting cases which can be analysed easily within the present framework. 
The first one corresponds to finite domain with a non-trivial boundary. In these situations, it is reasonable to require that the profile $\alpha$ approaches a global minimum of the energy-density at the boundary. In terms of the variable $H$ defined in Eq. (\ref{ans7}) this implies that $H$ must vanish on the boundary of the region one is considering. Due to the fact that, in the sector in which the non-linear superposition law is available, $H$ satisfies the Helmholtz equation, one arrives at the well-defined mathematical problem to solve the Helmholtz equation with vanishing Dirichlet boundary conditions. This problems is well analysed on mathematical textbooks and explicit solutions exist when the domain has enough symmetries. As it has been already discussed, the total energy is finite whenever the absolute value of the solution is strictly less than one\footnote{This can be always achieved since once one solves the Dirichlet problem one can multiply the solution by a small enough factor to satisfy the above requirement satisfying at the same time the same boundary conditions}.
The second case corresponds to infinite domains and so the mathematical problem is to solve the Helmholtz equation with vanishing Dirichlet boundary conditions in infinite domains. In these cases however, to require that the absolute value of the solution is strictly less than one is not enough to get a finite total energy: in the second case, the total energy diverges as for plane waves. Moreover, the time-dependence of the present periodic Skyrmions is similar to the usual behaviour of plane-waves. As it has been discussed in \cite{coleman} in the case of exact plane waves in Yang-Mills theory, in order to have well-behaved configurations it is enough to require that the energy-density is bounded and this happens in the present case as well. Actually, the present periodic multi-Skyrmions in infinite domains are more general than the exact multi-plane waves considered in the case of Yang-Mills theory, for instance, in \cite{devega}. The reason is that, both in \cite{coleman} and in \cite{devega}, in order to construct exact plane-waves of Yang-Mills theory in which the non-linear effects are present only configurations with flat wave-fronts were considered. On the other hand, in the present case, the spatial dependence of the profile of the periodic multi-Skyrmion (which determines the form of the wave-front) can be found by solving the Helmholtz equation and this allows to construct wave-front with non-trivial spatial geometries while keeping both the non-linear superposition law and the non-linear effects alive.

\section{Mapping between dual periodic Skyrmions}

In the previous section, it has been shown how to construct exact Skyrme
configurations which exhibit a non-linear superposition whose internal
orientation changes periodically in time while the Skyrmions profile $\alpha$
is static. Hence, it is natural to wonder whether it is possible to realize
also the "dual configurations" in which the profile $\alpha$ depends on time
while the internal orientation changes periodically along some space-like
direction. The equations of motion in both cases are very similar but there is
an important difference as it will be now shown.

Let us consider the following choices of the profile of the hedgehog $\alpha$
and of the function $\Theta$:%
\begin{equation}
\alpha=\alpha(t,x,y)  \ ,\ (\nabla\alpha)^{2}=-\left(
\partial_{t}\alpha\right)  ^{2}+(\partial_{x}%
\alpha)  ^{2}+(\partial_{y}%
\alpha)  ^{2} , \label{newnew1}%
\end{equation}
\
\begin{equation}
\Theta=L^{-1}z\  ,\ \ (\nabla\Theta)^{2}=L^{-2}\ , \label{newnew2}%
\end{equation}
where the flat Minkowski metric in Cartesian coordinates is used, $x$, $y$ and $z$ are the corresponding spatial coordinates and the constant $L$ has the dimension of length. The ansatz in Eqs. (\ref{newnew1}) and (\ref{newnew2}) (which is the natural
extension of the one in Eqs. (\ref{ans1}) and (\ref{ans2})) describes
Skyrmions with a time-dependent profile $\alpha$ which depends on two
space-like coordinates as well. The internal vector $\widehat{n}^{i}$ which
describes the orientation of the Skyrmion in the internal $SU(2)$ space
depends periodically on $z$.

With the above choice, the full Skyrme field equations reduce to%
\begin{equation}
0=\left(  1+\lambda L^{-2}\sin^{2}\alpha\right)  \square\alpha+\frac{\lambda
L^{-2}\left(  \nabla\alpha\right)  ^{2}}{2}\sin\left(  2\alpha\right)
-\frac{L^{-2}}{2}\sin\left(  2\alpha\right)  =0\ ,\ \ \square=-\partial_{t}%
^{2}+\partial_{x}^{2} +\partial_{y}^{2}. \label{newnew3}%
\end{equation}
In terms of the new variable $u$ defined as%
\[
\alpha=u+\frac{\pi}{2}\ ,
\]
Eq. (\ref{newnew3}) becomes%
\begin{align}
0  &  =\left(  1-\lambda\left(  \omega_{eff}\right)  ^{2}\sin^{2}u\right)
\square u+\frac{\left(  \omega_{eff}\right)  ^{2}}{2}\sin(2u)-\frac
{\lambda\left(  \omega_{eff}\right)  ^{2}\sin(2u)}{2}(\nabla u)^{2}%
\ ,\label{newnew4}\\
\left(  \omega_{eff}\right)  ^{2}  &  =\frac{L^{-2}}{1+\lambda L^{-2}}\ ,
\label{newnew5}%
\end{align}
where trivial trigonometric identities have been used. Hence, in terms of the
variable $u$, the equation is the same as equation (\ref{eom2}) analyzed in
the previous section with an effective frequency $\omega_{eff}$ given in Eq.
(\ref{newnew5}). In fact, it is now apparent the fundamental difference
between this case and the periodic Skyrmions described in the previous
section. One of the key observations of the previous section is that Eq.
(\ref{eom2}), when the critical condition in Eq. (\ref{freres}) is satisfied,
can be mapped into a linear equation. In the present case, the critical
condition corresponding to Eq. (\ref{freres}) would be%
\[
\lambda\left(  \omega_{eff}\right)  ^{2}=1\ ,
\]
so that, as it is evident from Eq. (\ref{newnew5}), the above condition cannot
be fulfilled for the ansatz in Eqs.
(\ref{newnew1}) and (\ref{newnew2}). This is the crucial difference between
the previous case (in which $\widehat{n}^{i}$ depends periodically on time)
and the one discussed in the this section. Although a sort of duality mapping
between the two cases can be constructed, the non-linear superposition law
only appears in when the profile $\alpha$ is static and the internal
orientation changes periodically in time.

\section{Conclusions}

In the present paper, exact configurations of the four-dimensional Skyrme
model have been constructed. Such configurations can be static and kink-like or time-periodic. The static configurations represent domain-walls: to the best of author's knowledge, these are the first exact domain-walls in the four-dimensional Skyrme theory. Within the class of the time-periodic configurations (periodic
Skyrmions), which can depend in a non-trivial way on all the space-like
coordinates, it is possible to disclose a remarkable phenomenon. For a special
value of the frequency, a non-linear superposition law arises which allows to
compose two (or more) periodic Skyrmions into a new one. The non-linear
superposition leads to the appearance of crystals of periodic Skyrmions in
which the peaks in the energy-densities of elementary Skyrmions are placed
according to patterns of negative interference. Such composite Skyrmions do
not disappear if the frequency is close enough to the resonance frequency
(although they will suffer small distortions). In a
non-linear theory such as the four-dimensional Skyrme model, these results are
very surprising and shed considerable new light on the interactions of Skyrmions.

\subsection*{Acknowledgements}

I warmly thank Hideki Maeda for many advises and I also thank very much
Francisco Correa, Hideki Maeda and Cedric Troessaert for carefully checking
the computations of this work. This work has been funded by the Fondecyt
grants 1120352. The Centro de Estudios Cient\'{\i}ficos (CECs) is funded by
the Chilean Government through the Centers of Excellence Base Financing
Program of Conicyt.


\begin{thebibliography}{99}                                                                                               %


\bibitem {skyrme}T. Skyrme, \textit{Proc. R. Soc. London} \textbf{A 260}, 127
(1961); \textit{Proc. R. Soc. London} \textbf{A 262}, 237 (1961);
\textit{Nucl. Phys.}\textbf{\ 31}, 556 (1962).


\bibitem {multis2}H. Weigel, \textit{Chiral Soliton Models for Baryons},
(Springer Lecture Notes 743)

\bibitem {manton}N.~Manton and P.~Sutcliffe, \textit{Topological Solitons},
(Cambridge University Press, Cambridge, 2007).


\bibitem {susy}David I. Olive and Peter C. West (Editors), \textit{Duality and
Supersymmetric Theories}, (Cambridge University Press, Cambridge, 1999).


\bibitem {bala2}A.P. Balachandran, H. Gomm, R.D. Sorkin, \textit{Nucl. Phys}.
\textbf{B 281} (1987), 573-612.

\bibitem {bala0}A.P. Balachandran, A. Barducci, F. Lizzi, V.G.J. Rodgers,\ A.
Stern, \textit{Phys. Rev. Lett.} \textbf{52 }(1984), 887.

\bibitem {bala1}A.P. Balachandran, F. Lizzi, V.G.J. Rodgers,\ A. Stern,
\textit{Nucl. Phys}. \textbf{B 256}, (1985), 525-556.

\bibitem {ANW}G. S. Adkins, C. R. Nappi, E. Witten, \textit{Nucl. Phys}.
\textbf{B 228} (1983), 552-566.

\bibitem {guada}E. Guadagnini, \textit{Nucl. Phys}. \textbf{B 236} (1984), 35-47.

\bibitem {astroSkyrme}H. Pais, J. R. Stone, \textit{Phys. Rev. Lett}.
\textbf{109} (2012), 151101.

\bibitem {useful1}Al Khawaja, Usama; Stoof, Henk,. \textit{Nature}
\textbf{411} (2001): 918--20.

\bibitem {useful2}J.-I. Fukuda, S. Zumer, \textit{Nature Communications}
\textbf{2} (2011), 246.

\bibitem {useful3}Christian Pfleiderer, Achim Rosch, \textit{Nature}\textbf{
465}, 880--881 (2010).

\bibitem {useful4}S. M\"{u}hlbauer, B. Binz, F. Jonietz, C. Pfleiderer, A.
Rosch, A. Neubauer, R. Georgii, P. B\"{o}ni, \textit{Science} \textbf{323}
(2009), pp. 915-919

\bibitem {useful5}F. Jonietz, S. M\"{u}hlbauer, C. Pfleiderer, A. Neubauer, W.
M\"{u}nzer, A. Bauer, T. Adams, R. Georgii, P. B\"{o}ni, R. A. Duine, K.
Everschor, M. Garst, A. Rosch, \textit{Science} \textbf{330} (2010), pp 1648-1651\ 

\bibitem {useful6}S. Seki, X. Z. Yu, S. Ishiwata, Y. Tokura, \textit{Science}
\textbf{336} (2012), pp. 198-201.

\bibitem {useful7.1}U. K. Roessler, A. N. Bogdanov, C. Pfleiderer,
\textit{Nature} \textbf{442}, 797 (2006).

\bibitem {useful7}A. N. Bogdanov D. A. Yablonsky, Sov. Phys. JETP 68, 101 (1989).

\bibitem {useful8}A. Bogdanov and A. Hubert, J. Magn. Magn. Mater. 138, 255
(1994); 195, 182 (1999).


\bibitem {moduli}N. S. Manton, \textit{Phys. Lett}.\textbf{ 110B}, 54 (1982).


\bibitem {skyrmonopole}Y. M. Cho, \textit{Phys. Rev. Lett.} \textbf{87
}(2001), 252001.

\bibitem {knotYM}L. Faddeev, A. Niemi, \textit{Nature} \textbf{387}, 58 (1997).

\bibitem {example1}M. Hirayama, J. Yamashita, \textit{Phys. Rev}.\textbf{ D
66}, 105019 (2002).

\bibitem {sut1}R. A. Battye, P. M. Sutcliffe, \textit{Phys. Rev. Lett}.
\textbf{79}, 363 (1997); \textit{Phys.Lett. \ }\textbf{B 391} (1997) 150-156.

\bibitem {rational}C. J. Houghton, N. S. Manton, P. M. Sutcliffe,
\textit{Nucl. Phys}. \textbf{B 510}, 507 (1998).

\bibitem {simple1}B.M.A.G. Piette, B.J. Schoers and W.J. Zakrzewski,
\textit{Z. Phys}. \textbf{C 65} (1995) 165; B.M.A.G. Piette, B.J. Schoers and
W.J. Zakrzewski, \textit{Nucl. Phys}. \textbf{B 439} (1995) 205.

\bibitem {simple2}R.A. Leese, M. Peyrard, W.J. Zakrzewski,
\textit{Nonlinearity} \textbf{3 }(1990) 773; B.M.A.G. Piette, W.J. Zakrzewski,
Chaos, \textit{Solitons and Fractals} \textbf{5} (1995) 2495; P.M. Sutcliffe,
\textit{Nonlinearity }\textbf{4} (1991), 1109.

\bibitem {simple3}T. Weidig, \textit{Nonlinearity} \textbf{12} (1999) 1489.

\bibitem {simple4}P. Eslami, M. Sarbishaei and W.J. Zakrzewski,
\textit{Nonlinearity} \textbf{13} (2000) 1867.

\bibitem {simple5}A.E. Kudryavtsev, B. Piette, W.J. Zakrzewski, \textit{Eur.
Phys}. \textit{J}. \textbf{C1}, 333 (1998).

\bibitem {simple6}C. Adam, J. Sanchez-Guillen, A. Wereszczynski, \textit{Phys.
Lett}. \textbf{B 691} (2010), 105; \textit{Phys. Rev}. \textbf{D 82} (2010), 085015.

\bibitem {canfora}F. Canfora, P.~Salgado-Rebolledo, \textit{Phys. Rev.}
\textbf{D 87}, 045023 (2013).

\bibitem {canfora2}F. Canfora, H.~Maeda, \textit{Phys. Rev.} \textbf{D 87},
084049 (2013).

\bibitem {sut2}P. M. Sutcliffe, \textit{Nucl. Phys}. \textbf{B 431} (1994), 97.

\bibitem {vilenkin} A. Vilenkin, E. P. S. Shellard, \textit{Cosmic Strings and Other Topological Defects}, Cambridge University Press (2000).

\bibitem {berger}M. Berger, \textit{Non-linearity and Functional Analysis},
Academic press, 1977.

\bibitem {coleman} S. Coleman, \textit{Phys. Lett.} \textbf{70 B},
59 (1977).

\bibitem {devega} H. J. de Vega, \textit{Comm. Math. Phys.} \textbf{116},
659 (1988).


\end{thebibliography}
\end{document}